\documentclass[manuscript]{acmart}
\AtBeginDocument{%
  \providecommand\BibTeX{{%
    \normalfont B\kern-0.5em{\scshape i\kern-0.25em b}\kern-0.8em\TeX}}}

\newcommand{\tmpout}[1]{}

\setcopyright{none}

\usepackage{hyperref}
\usepackage{caption}
\usepackage{subcaption}

\begin{document}

\title{AI-Resilient Interfaces [Working Draft]} %:
%Designing for AI Safety, \\Utility, \& Usability during Open-Ended %Tasks \\ or \\
%AI-Resilient Interfaces: 
%Improving AI Safety \&  Utility by Making AI's Choices Easier to Notice, Judge Well, \& Recover From}

\author{Elena L. Glassman}
\authornote{Glassman proposed the core initial idea, which Glassman and Kummerfeld then developed together, with input from Glassman's lab.
The Introduction, Search Example, and Relationship with Trust and Design Principles, were primarily written by Glassman.}
\email{glassman@seas.harvard.edu}
\orcid{https://orcid.org/0000-0001-5178-3496}
\affiliation{%
  \institution{Harvard University}
  \streetaddress{John A. Paulson School of Engineering \& Applied Sciences}
  \city{Boston}
  \state{Massachusetts}
  \country{USA}
  \postcode{02134}
}

\author{Ziwei Gu}
\authornote{Advised by Glassman and Kummerfeld, Gu implemented the first AI-resilient alternative to automated text summarization, Grammar-Preserving Text Saliency Modulation~\cite{gptsm} and suggested additions to this paper.}
\email{ziweigu@g.harvard.edu}
\orcid{https://orcid.org/0000-0001-9044-2651}
\affiliation{%
  \institution{Harvard University}
  \streetaddress{John A. Paulson School of Engineering \& Applied Sciences}
  \city{Boston}
  \state{Massachusetts}
  \country{USA}
  \postcode{02134}
}

\author{Jonathan K. Kummerfeld}
\authornote{The Corpus Example and Audits were primarily written by Kummerfeld.\\
All other sections were jointly written by Glassman and Kummerfeld;
both authors edited and gave feedback on all sections.}
\email{jonathan.kummerfeld@sydney.edu.au}
\orcid{https://orcid.org/0000-0001-5030-3016}
\affiliation{%
  \institution{The University of Sydney}
  \streetaddress{School of Computer Science Building (J12)}
  \city{Darlington}
  \state{New South Wales}
  \country{Australia}
  \postcode{2006}
}

\renewcommand{\shortauthors}{Glassman, Gu, and Kummerfeld}

\begin{abstract}
%We need interfaces that help people notice and better evaluate AI choices, and be resilient to the choices that are not right, or not right for them. 
AI is powerful, but it can make choices that result in objective errors, contextually inappropriate outputs, and disliked options. 
We need \emph{AI-resilient interfaces} that help people be resilient to the AI choices that are not right, or not right for them. To support this goal, interfaces need to help users \textit{notice} and \textit{have the context to appropriately judge those AI choices}.  
Existing human-AI interaction guidelines recommend efficient user dismissal, modification, or otherwise efficient recovery from AI choices that a user does not like. However, in order to recover from AI choices, the user must notice them first. This can be difficult. For example, when generating summaries of long documents, a system's exclusion of a detail that is critically important to the user is hard for the user to notice. That detail can be hiding in a wall of text in the original document, and the existence of a summary may tempt the user not to read the original document as carefully.
Once noticed, judging AI choices well can also be challenging. 
The interface may provide very little information that contextualizes the choices,
and the user may fall back on assumptions when deciding whether to dismiss, modify, or otherwise recover from an AI choice. 
Building on prior work, this paper defines key aspects of AI-resilient interfaces, illustrated with examples. Designing interfaces for increased AI-resilience of users will improve AI safety, usability, and utility. This is especially critical where AI-powered systems are used for context- and preference-dominated open-ended AI-assisted tasks, like ideating, summarizing, searching, sensemaking, and the reading and writing of text or code.
%The user may have very little information or information scent to guide their evaluation of an AI choice; they may fall back on assumptions to guide them in their judgement. 
%That judgement is what drives how the user wields any previously recommended affordances for recovering from an AI choice, and we argue that many critical current interfaces poorly support this judgement and its precondition, noticing.

%

%
\end{abstract}





%   \includegraphics[width=\textwidth]{figures/Human-Computer_Communication_Figure.jpg}
%   \caption{AI-resilient interfaces concern the support for noticing and (re)examining AI outputs with as much common ground as is possible to enable the highest quality user judgements about whether or not AI outputs are appropriate for their context. Figure adapted from \cite{hcc_glassman}.}
%   %\Description{Human and computer... TBD}
%   \label{fig:hcc_figure}

%\received{date}
%\received[revised]{date}
%\received[accepted]{date}

\maketitle

% \todo{Sensible AI paper - Elena}

% \todo{Resolve other notes in Related Work}
% %

% \todo{discuss resiliency fails of various AI actions: extract, rank, filter, organize}

%\todo{finish adding additional search example -- Elena}
%\todo{add revised new talk abstract to beginning? - Elena}

%\todo{add reference to iceberg sensemaking paper when discussing context to make decisions}

%\todo{add language and examples from ETH talk}

%\todo{add interleaved PDC}

\section{Introduction}

AI and other forms of automation are powerful, but their computational \emph{choices} can result in objective errors as well as contextually inappropriate or subjectively insufficient outputs.
For example, legal technology companies are piloting AI-written summaries for judges in lieu of those written by paralegals. A summary is a lossy representation that is useful if it appropriately frames any additional details in the original case document, i.e., preserving factors that are contextually or personally relevant to the reader, given the reader's particular knowledge, context, preferences, values, and task. In the case of a judge, the details that the judge believes are crucial to contextualizing the offense might be left out. 
%AI that is generating a summary in place of a paralegal 
The AI may “choose” not to include such relevant factors (omission), introduce statistically correlated---and therefore plausible and harder to trigger suspicion---info (hallucination), and
subtly or significantly change the semantic meaning through rewording or leaving some contextualizing information out (misrepresentation)~\cite{gptsm}. %Automated summarization methods can introduce multiple types of errors: omission, hallucination, and misrepresentation~\cite{gptsm}. 

%Omission can show up, for example, when these methods judge some details as insufficiently relevant and omit them when they are actually crucial to the reader, given the reader's particular knowledge, context, preferences, values, and task. In the case of a judge, the details that the judge believes are crucial to contextualizing the offense might be left out.

\textbf{We need \emph{AI-resilient interfaces} that help people be resilient to the AI choices that are not right, or not right for them.} To support this goal, interfaces need to help users \textit{notice} and \textit{have the context to appropriately judge those AI choices}. This will improve AI safety, usability, and utility.
In the example of automated summarization of long documents or document corpora, the reader has to read and/or remember both the summary and the source document(s) to notice and accurately judge whether the AI summary includes any omissions, hallucinations, and misrepresentations relevant to their task. 
Catching these issues is onerous, and due to memory constraints, error-prone; it introduces extra work for the user if they want to be resilient to them. It therefore does not meet our design goals for AI-resilient interfaces. One could introduce additional AI to attempt to identify and call out these omissions, hallucinations, and misrepresentations, but that does not eliminate the need for the reader check the additional AI feature's choices; it may help the user catch more issues between the original AI-generated summary and its summarized document(s), but it can still introduce a new class of objective errors and contextually inappropriate or subjectively insufficient choices.

\paragraph{Noticing} Existing human-AI interaction guidelines, e.g.,~\citet{amershi2019guidelines}, often recommend including affordances for efficient user dismissal, modification, or otherwise efficient recovery from AI choices that a user does not like. However, in order to recover from AI choices, the user must first notice them, which is not always trivial, like the challenge of noticing a crucial detail that an AI left out. There are many ways, e.g., inattentional blindness~\cite{inattentional}, in which users may fail to notice consequential AI choices when examining system outputs.

\paragraph{Judging (well)} Once noticed, judging AI choices well is also not trivial. Humans do not have computers' tireless capability to consume all the relevant data in its original form that \textit{would} provide more of the necessary context to judge an AI choice. The interface may provide very little contextual information 
and the user may fall---consciously or unconsciously, confidently or hesitantly---back on assumptions to guide them in their judgment. If the interface includes an AI system's explanations of its own behavior, there can still be insufficient context provided for the human to recognize when to overrule it, if the user is sufficiently engaged with the explanation at all. 
If the interface includes AI estimates of its own uncertainty, those estimates can be poorly calibrated and do not protect users when the AI is confidently wrong. 
Likewise, the AI can also not be relied upon to make its own judgements about what AI choices are or are not part of the set that the user would need to see and understand to be resilient to the AI's choices, because again the AI can be confidently wrong. 
The user's assessment of the situation, however well-informed or flawed, is what drives how the user wields the recommended affordances to dismiss, modify, or otherwise recover from an AI choice. 

In other words, the gulf of evaluation~\cite{gulfbook} when using AI-powered features is often very large and impractical or nearly impossible to traverse. Ideally, AI-resilient systems should narrow or nearly eliminate this gulf with the specific goal that, when the AI is making choices that are wrong (or ``just'' wrong for that particular user), the user is able to notice and accurately judge the quality of those choices without needing to significantly go out of their way to do so. (If an interface allows a user to recognize more wrong or suboptimal AI choices than existing interfaces, but it still requires significant additional user effort, one could say that the interface is allowing the user to be more AI-resilient, which is obviously good, while not fulfilling the design ideal.)  
%As a result, we hope AI-resilient interfaces will 
%will allow users' abilities to be enhanced whenever the AI is not wrong, and 
%introduce minimal extra cognitive or physical work when the AI is wrong, such that users are either accelerated or unharmed.

%n objective error, subjectiv helping the user be just as or better off than 
%continue on with their task just as or better than before without adding new labor. 
%One could consider this critique of existing systems as the \textit{``gulf of evaluation~\cite{gulfbook} on steroids''}
%, in hyperrealistic detail, 
%for the specific context of human-AI interaction. 

AI-resilient interfaces are especially important for the utility and usability of AI in context- and preference-dominated and/or open-ended AI-assisted tasks, like ideating, summarizing, searching, sensemaking, and the reading and writing of text or code. For example, when writing, there is no objectively correct next sentence, only a large design space of possible thoughts and their concrete instantiations. Only the writer knows their current, evolving beliefs, what communicative goals they have, and what literal text they now want.

In this paper, using examples, we argue that many critical current interfaces poorly support this process of noticing and judging.
%We need interfaces that help people notice and better judge AI choices, in order for these users to be resilient to the AI choices that are not right, or not right for them. One could consider this focus as the gulf of evaluation~\cite{gulfbook} on steroids, in hyperrealistic detail, for the specific context of human-AI interaction. 
And designing specifically for these goals is not trivial. 
%An AI may implicitly make prohibitively many potentially consequential choices in the process of generating an output. Are only a subset of these AI choices and/or their downstream impacts necessary to make legible in an interface to enable the user to be resilient to AI choices they disagree with? How can these AI choices and/or impacts be reified in the interface without overwhelming the user? At what point are any original perceived value of AI to the user nullified, e.g., by causing choice paralysis rather than clarity? If all the relevant AI choices and/or impacts cannot be reified in the interface, how does one reveal a subset without creating a situation where an omission fundamentally distorts the user's interpretation and subsequent judgement of an AI choice's correctness, appropriateness, or utility, as illustrated by an example from Google Search in Section~\ref{sec:twins}? 
Given the challenges of developing AI-resilient interfaces as defined above, it may seem like a fools' errand to spend so much time defining this characteristic that perhaps no interface could ever instantiate. However, we have one example that we believe fully satisfies this specification: Grammar-Preserving Text Saliency Modulation~\cite{gptsm}, an alternative to AI-generated document summaries that allows the user to read the original document more quickly, with just as much comprehension, with AI suggestions for where to focus reified in the (always legible) word-by-word saliency.
Given this existence proof, this paper picks up where that previous paper left off: ``generalizing this notion of AI-resiliency to additional tasks and domains.''

\section{Motivating Examples} \label{sec:twins}

The critical importance of noticing and judging is illustrated in the following three examples of how \textit{not} designing for noticing or judging can create usability, utility, and safety issues for users, even in mundane, pervasive interface types. The first two examples take place in the context of AI-assisted question answering using a search engine. The third example describes the corresponding noticing and judging challenges users face when using AI-assisted document clustering for analysis.

\subsection{AI-assisted Search}
\subsubsection{Twins as a function of maternal age}\label{sec:twins}

\begin{figure}[b]
    \centering
\includegraphics[width=0.8\textwidth]{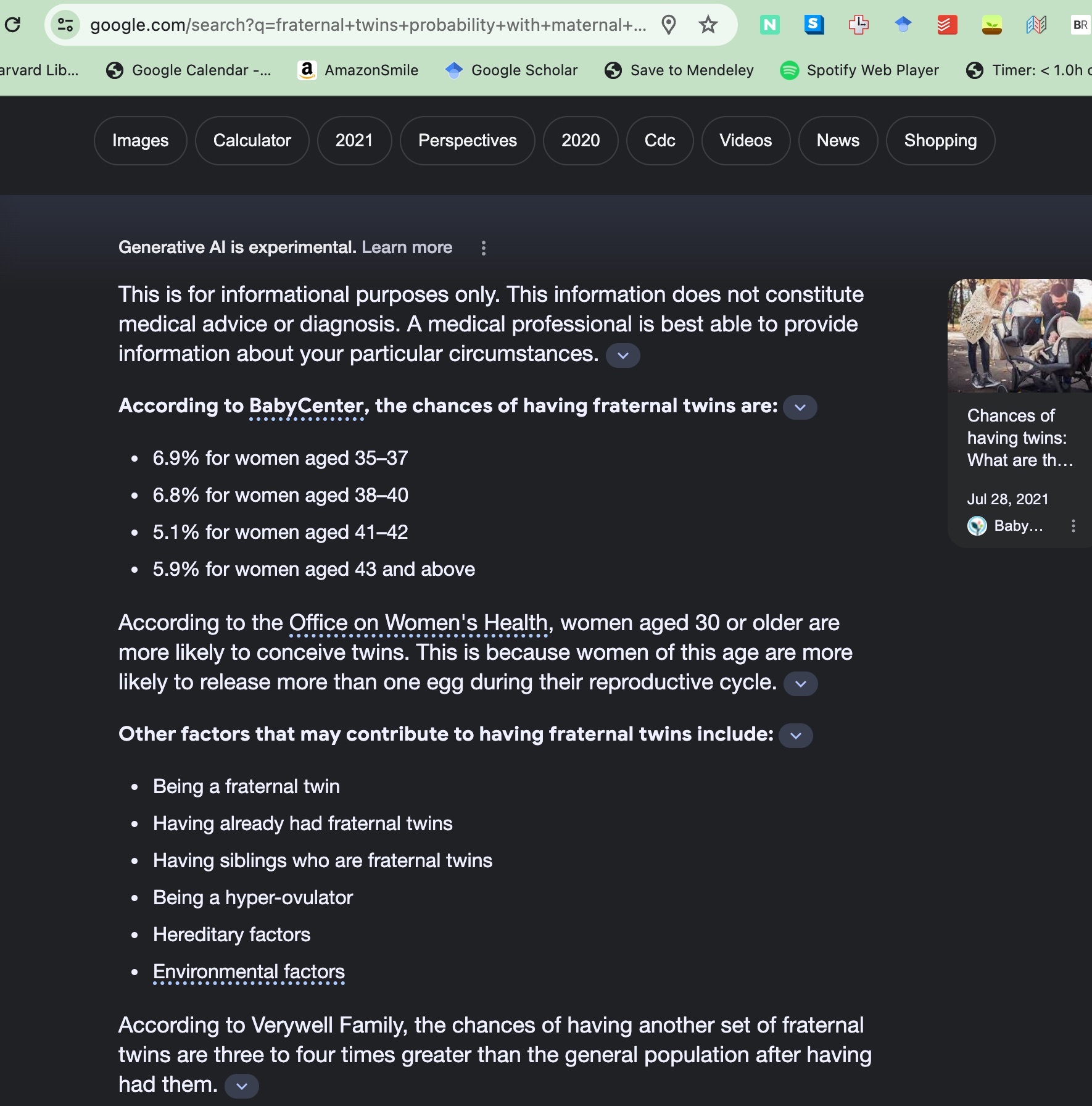}
    \caption{Generative AI's output in response to the query \texttt{fraternal twins probability with maternal age}.}
    \label{fig:twinsGenAI}
\end{figure}

Recently, the first author tried to look up the probability of having twins as a function of age, as they vaguely recalled that older mothers were more likely to have multiples. They searched Google for \texttt{fraternal twins probability with maternal age}.
The first answer on the search results page was attributed to Google's experimental generative AI, shown in Figure~\ref{fig:twinsGenAI}. The AI-generated text named a source's domain name (\texttt{BabyCenter.com}), generated some introductory text restating some of the question being answered, i.e., \texttt{the chances of having fraternal twins are}, and then listed percentages as a function of two-year age ranges.\footnote{Interestingly, both the original query, \texttt{fraternal twins probability with maternal age}, and the introductory generated text, \texttt{the chances of having fraternal twins are}, make no mention of maternal age, but the answer only includes the relationship between twins and age for those of "advanced maternal age"---35 or older. There are multiple possible reasons for this omission which are difficult or impossible to generate and distinguish between as a user of this black-box system; for example, perhaps the system leveraged its information about the searcher who was signed into their profile at the time and is already of "advanced maternal age" or perhaps this simply propagated the omissions of the cited source material.} 

The generative AI answer continued, moving on to another source which presumably independently confirmed that \texttt{women aged 30 or older are more likely to conceive twins}, complete with a reasonable sounding reason, followed by a list of many additional factors associated with having twins. While the searcher found the specific rates of 6.9\%, 6.8\%, 5.1\%, and 5.9\% for women in various age brackets (all over 35) surprisingly high, the response appeared well cited, consistent, and comprehensive, and confirmed the searcher's suspicions.

Further down the page, another intelligent feature called ``People Also Ask'' listed a semantically consistent rephrasing of the searcher's query as a grammatically correct sentence: \texttt{How common are twins by maternal age?} with an answer very similar to the generative AI's answer (Figure~\ref{fig:twinsExtracted}). The automatically generated answer was much shorter: just a list of percentages as a function of maternal age, with a date (presumably the date of retrieval) followed by the specific source page's title and domain name. The percentages were in the same range, with slightly different syntax and formatting. Given these superficial differences, a searcher may or may not notice that the numbers were actually exactly the same as those in the generative AI's answer, as was the domain name of the source. But this at a glance illusion of independent confirmation is not the primary problem we have with this interface, as revealed by what happened next:

\begin{figure}
    \centering
    \includegraphics[width=0.8\textwidth]{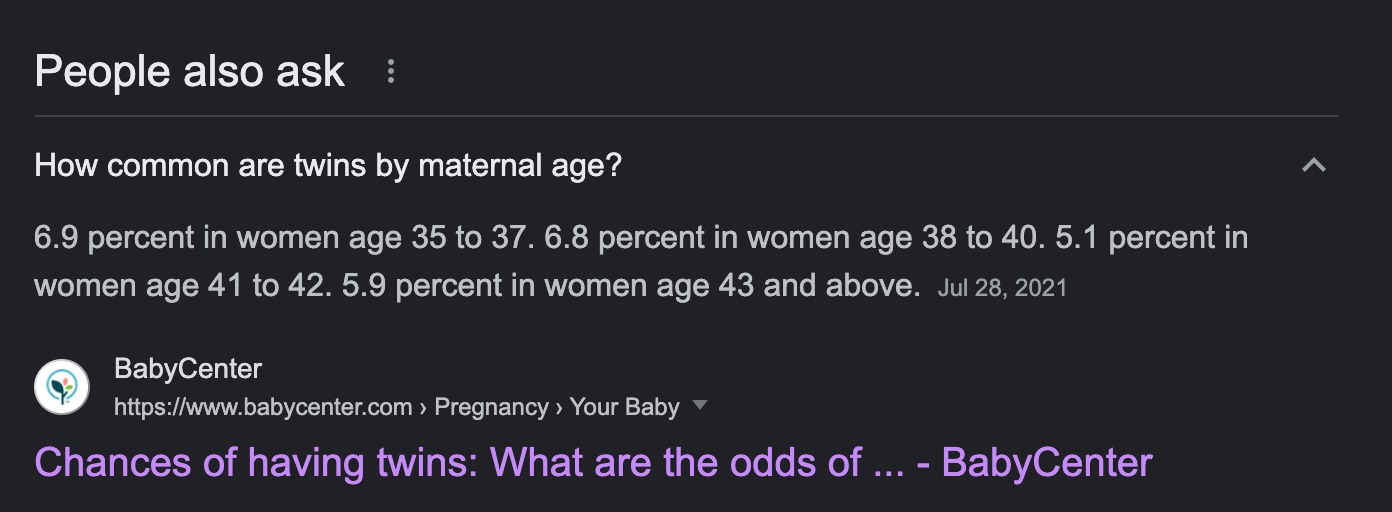}
    \caption{Another automated answer, which is a selected quote from the referenced page}
    \label{fig:twinsExtracted}
\end{figure}

\begin{figure}
    \centering
\includegraphics[width=0.8\textwidth]{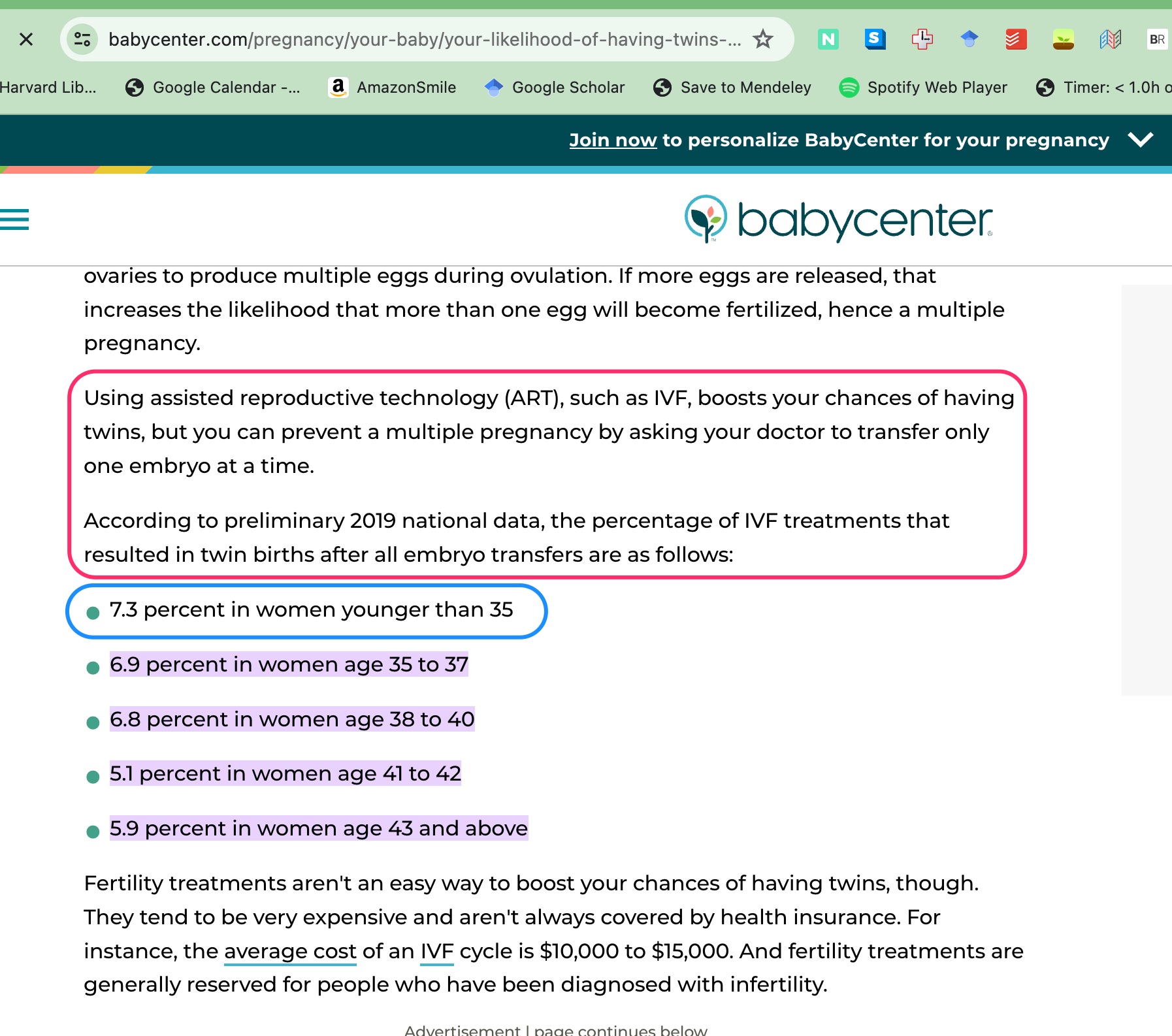}
    \caption{The actual referenced page~\cite{babycenter}. The blue and pink boxes highlight two distinct, critical pieces of information present on the original page that were automatically omitted from the extracted or generated answers on the original Google search page, misleading the searcher.}
    \label{fig:groundtruth}
\end{figure}

The searcher then chose to click on the specific page cited for these numbers in the second generated answer (Figure~\ref{fig:twinsExtracted}). The searcher's browser opened the page in a new window, which then automatically scrolled to and highlighted the text related to the search query (Figure~\ref{fig:groundtruth}, purple highlighting). This source page reveals several key facts: 

\begin{enumerate}
    \item The answer in the ``People Also Ask" feature is a contiguous excerpt from the linked page. (This, in and of itself, is not a problem.)
    \item This extracted answer omits the data point prior to the extracted data points for women younger than 35, which is even higher than the twin rate for women in the age brackets over 35 (Figure~\ref{fig:groundtruth}, blue box content). This omission allows the searcher to \textit{fill in the missing younger women's twin rate with their own expectations}, which are likely lower than these high quoted twin rates, since most people do not meet twins as frequently as these quoted rates would imply. \textit{The searcher may not even notice that they are doing this.}
    \item All this data actually only describes the twin rate for those using assisted reproductive technology (ART), where the practice of transferring multiple fertilized embryos in a single cycle is not uncommon (Figure~\ref{fig:groundtruth}, pink box content). This twinning rate has nothing to do with natural twinning rates that do in fact rise with maternal age and everything to do with irrelevant factors, i.e., common ART practices, and as evidenced by the omitted data point for younger women, does not increase with age.
\end{enumerate}

In other words, while it might still be true that unassisted older women have twins more often than younger women due to aging processes, the twin rate is not necessarily at the high rates observed in ART clinics that the AI quoted. 

\subsubsection{Folate supplementation during pregnancy}
\begin{figure}
     \centering
     \begin{subfigure}[b]{0.3\textwidth}
         \centering
         \includegraphics[width=\textwidth]{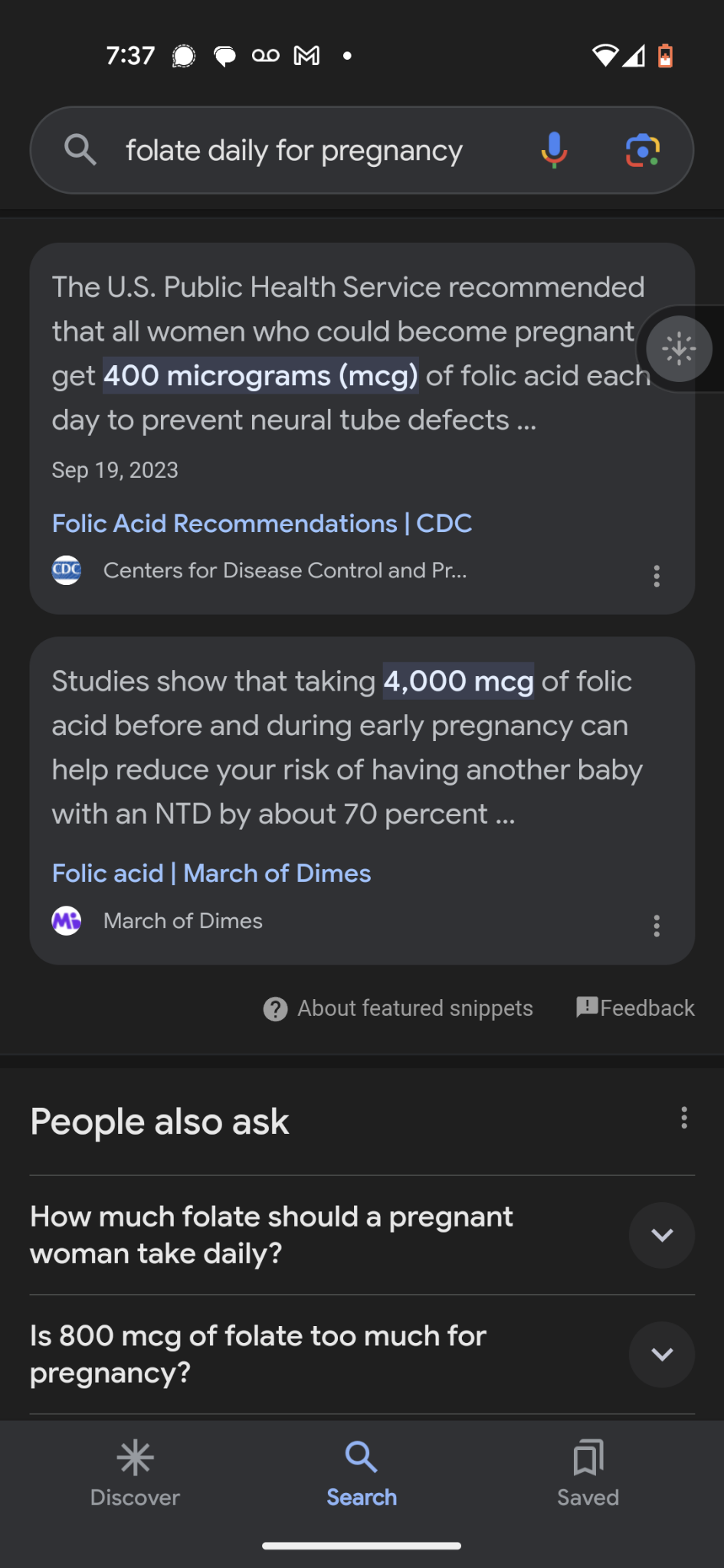}
         \caption{Google's extracted quotes as answers to the query \texttt{folate daily for pregnancy}. The extracted answers from two different sources differ by 10x.}
         \label{fig:y equals x}
     \end{subfigure}
     \hfill
     \begin{subfigure}[b]{0.3\textwidth}
         \centering
         \includegraphics[width=\textwidth]{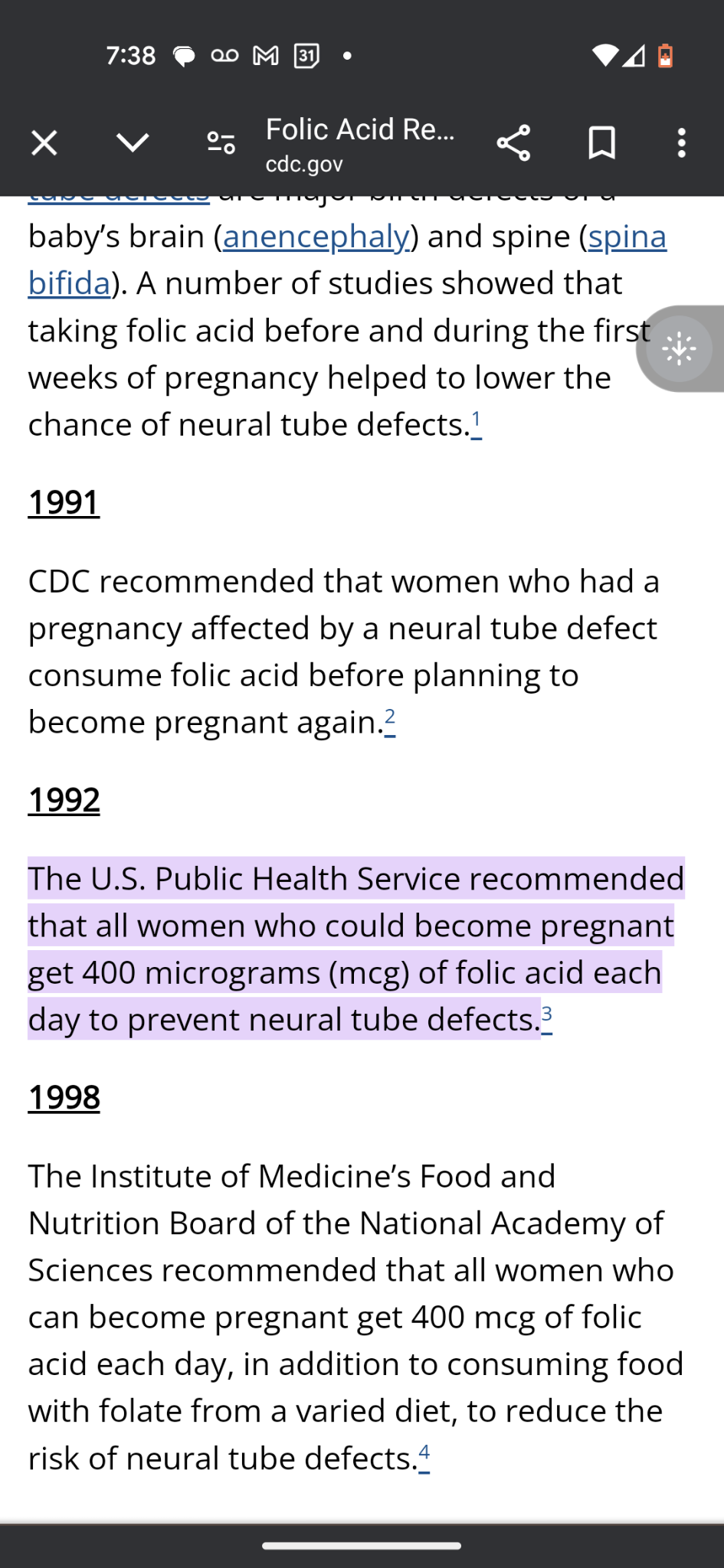}
         \caption{The first extracted quote in context. Is 400 mcg still the recommended dose? We only know from the local context of this quote that it was in 1992.}
         \label{fig:three sin x}
     \end{subfigure}
     \hfill
     \begin{subfigure}[b]{0.3\textwidth}
         \centering
         \includegraphics[width=\textwidth]{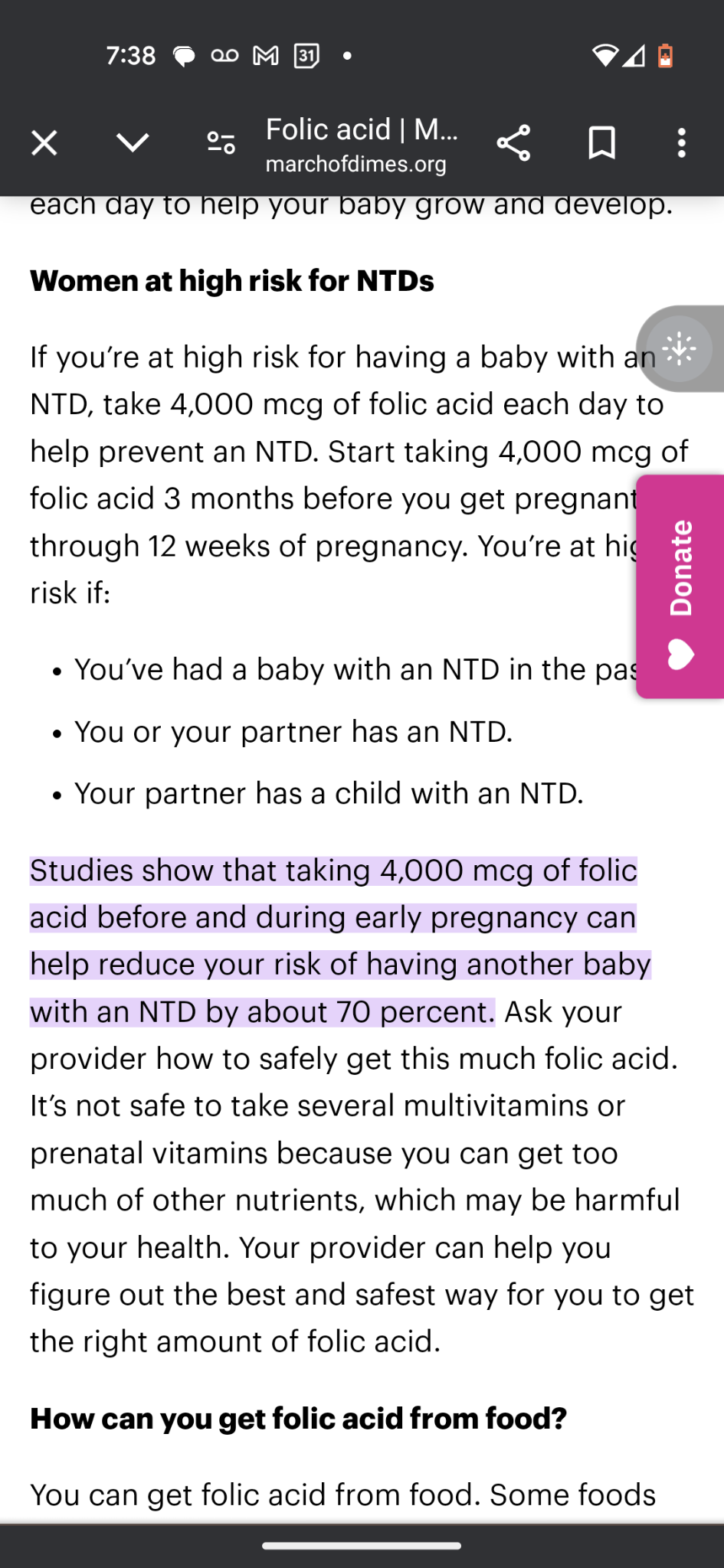}
         \caption{The second extracted quote in context. Turns out this recommendation is only for pregnancies at higher risk for neural tube defects.}
         \label{fig:five over x}
     \end{subfigure}
        \caption{An example of answers to the query \texttt{folate daily for pregnancy} as AI-extracted quotes initially provided with only the page title and organization name as context. The significant discrepancy in answers may be sufficient information scent for the user to take the necessary follow-up actions to be resilient to any of these AI choices that are not right for them.}
        \label{fig:folate_all}
\end{figure}

Shown in Figure~\ref{fig:folate_all}, the 10-fold magnitude difference in quoted folate dosage between the two AI-extracted answers from two different sources was large enough to be noticeable and considered significant enough by the first author to warrant gathering additional information. She opened up each resource, shown in (b) and (c) respectively, to better understand the discrepancy, and discovered that each answer was implicitly answering a slightly different question than she had originally intended to ask. That said, the diversity of answer revealed to her that there were questions she had not thought of ask---unknown unknowns---i.e., that some people are designated higher risk of gestating a fetus with a neural tube defect, and the recommended folate intake is different.

\paragraph{Summary}
Both the generative AI-produced and automatically extracted answers left out context that the system incorrectly judged to be insufficiently relevant. As a result, the system presented correct information without the context necessary to interpret it correctly, and did so with no information scent about the complexity or context abstracted away (appropriately or inappropriately) to generate that answer. There was nothing in the page of search results and generated answers for the searcher to \textit{notice}---let alone judge well the appropriateness of excluding---the missing context, and, when the extracted answers reinforced rather than contradicted each other, there was no information scent that signaled that the searcher might want to investigate further on the original source page, aside from the existence of a link to it. 
In these examples, to both (1) be correctly informed rather than misled and (2) have the context necessary to judge the AI output well, the searcher needed to have the curiousity to follow a source link and then notice that omitted text fundamentally changed how that quoted data should be interpreted.

To complicate matters, given inattentional blindness~\cite{inattentional}, a user being in a position to theoretically notice does not guarantee that they will indeed notice, and asking the system to draw users' attention to its own mistakes puts users \textit{again} at the mercy of its mistakes. (The system can be confidently wrong about both its chosen output and any ``meta cognition'' about its output.) In this case, the automated highlighting of the specific source material within the source page at least helped the searcher in this scenario to notice the relevant omitted context nearby. There may have been other relevant context on the rest of the page which the searcher never noticed; exhaustively noting all relevant omitted context would require reading the entire original page.

\subsection{AI-assisted Document Corpus Exploration}

In a range of settings, people wish to understand large collections of documents, e.g., finding trends on social media, understanding patterns in news, and analyzing survey responses.
Understanding a corpus involves identifying patterns, relationships, and groupings across the documents (where `documents' could be anything from single words to entire books). 
As well as the patterns of commonality, there is also value in identifying group boundaries and distinctions, as well as heads and long contrasting tails of power-law distributions, and outliers.

An interactive scatter plot is a standard method in this exploratory analysis.
Each point in the plot represents a document, and the spatial relationships between dots is determined by mathematical similarity functions applied to the documents' representation.
Color is often used to encode document meta data or computed clusters.
Hovering over a point shows the content of the document or other information.
Sometimes denser groupings are labeled automatically with a short descriptive label that captures something that is more prevalent among the dots in the neighborhood than elsewhere.
More sophisticated interaction methods have also been explored, such as a lasso tool that allows users to select and view information about a selected subset of dots~\cite{explainAndTest}.

Users may hope that these plots can both confirm expected relationships and reveal unexpected relationships.
In both cases, this hope is based on the ability to see a visual encoding of the entire document collection at once with semantic relationships represented spatially.
This is particularly beneficial when the document collection is inconveniently or even prohibitively large for users to read through in its entirety. 
Producing this standard plot requires multiple computational processes, which may or may not capture or preserve relationships that the user currently cares about---or would care about if they knew about them.\footnote{First, a model produces a high-dimensional vector representation of each document, which may or may not capture the aspects of that document that the user currently cares or would care about.
Optionally, second, an unsupervised clustering method identifies groups of vectors that are similar to each other, using a definition of similarity that may or may not capture relationships between documents that the user currently cares or would care about.
And third, a dimensionality reduction algorithm is used to convert the high-dimensional vectors into a 2D space, which may or may not preserve, in the final 2D spatial mapping, the relationships currently cares or would care about.}

The backend computational processes are making decisions for the user that are hard to meaningfully \emph{notice} in the standard interface, because the documents are represented \emph{as dots}. Nothing about that dot reveals the underlying text in the document that define between-document relationships---all the possible relationships, which is a spectrum from those that are computationally prioritized to those that the computer is functionally ``blind'' to. 
The user may be aware of the computational steps involved in generating the scatter plot, but the interface does not provide an effective way to notice (1) the consequences of particular distance metrics and/or thresholds shaping the process or (2) whether the outcome of this process has placed points---that they would want to be spatially close---far away or placed points close to each other that the user would want far away from each other. 
Finding such pairs of points amounts to the matching (memory) game,\footnote{The Matching Game or Memory Game is typically played with a 2D array of upside-down cards where players have to identify matching cards by flipping only one card right-side-up at a time to reveal its identity} where flipping a card to reveal its contents is akin to hovering over a document dot. Descriptive automated short labels over clusters of dots choose to call out a single commonality for the user, and even then, users would need to manually spot-check documents associated with various dots near and far from the label to attempt judge the quality of the label, and trust that they can accurately generalize from those samples to infer an overall pattern. There's no information scent for finding exceptions to the label's suggested pattern within the neighborhood or where and how one neighborhood transitions into another. The user may prioritize examining dots that are spatial outliers or otherwise spatially remarkable, but these spatial deviations are a function of the embedding and, as a result, may or may not reflect deviations from the rest of the corpus along aspects that are relevant to the user's task, context, and preferences. Exhaustive search and a superhuman working memory is necessary to fully take in what the embedding has actually done with the documents, let alone notice discrepancies along aspects they realize---as they examine documents---or already knew they care about.

One response to these issues could be that we need better algorithms for embedding generation, clustering, and dimensionality reduction.
While improvements to the AI components could help, they will not solve the issues discussed above because those issues are fundamentally about the way the visualization communicates the AI output to the user.
This lack of interface support for fully seeing and understanding the AI choices makes it much harder for the user to see when the AI has missed an aspect the user values.
These algorithms also all have knobs that can be adjusted to shape their behavior, but it is difficult for the user to know which knob to adjust (gulf of execution~\cite{gulfbook})---if they even see enough text to realize such adjustment is needed (gulf of evaluation~\cite{gulfbook}).
An AI-resilient method of visualizing documents sets and clusters should empower users both to look inside and across clusters, which in turn allows them to identify how the AI decisions do or do not suit their needs.

\paragraph{Summary} Not prioritizing user \textit{noticing} and \textit{judging}---or not supporting it at all---can lead to AI-powered interfaces misleading users or providing less utility than they are capable of.

\section{Design Challenges}

\subsection{There often is no ``best'' AI choice.}
AI choices are typically defined in terms of and evaluated relative to some ``objective'' ground truth, but in many contexts where AI-powered interfaces are now deployed, the user's context, goals, opinions, values, preferences and risk tolerances dominate, rather than ``objective'' notions of accuracy. These aspects of the user's situation 
%\todo{relate to situated cognition and Tamara's Iceberg interpretivist language} 
may be partially or completely unobservable. No matter how sophisticated the affordances are for the user to externalize the unobservable parts of their context, e.g., internal goals and relevant context, they would need to make the effort to identify and express them. In this respect, only the user is capable of determining whether an AI choice is wrong or not good enough in their eyes for their situation.\footnote{The ``best'' AI choice could be (1) ``reading the user's mind'' (which is, for the foreseeable future, an impossible task) in order to provide what what the user currently wants, (2) providing a spectrum of options that may help the user recognize what they already want (or what they now \emph{realize} they want), or (3) not giving what the user wants to explicitly challenge them and perhaps drive the human-AI team to a location that the human, when reflecting after the fact, realizes is a better outcome than they had originally envisioned.}

\subsection{Unnoticed (AI) choices}
\label{sec:unnoticed}
Our recognition of AI choices 
is not guaranteed, or even possible, in many interfaces. Sometimes those choices are literally hidden, 
and sometimes they are hidden in plain sight. 

\paragraph{1. Invisible Choices}
The user cannot see AI choices when the interface silently hides information based on that choice; AI choices users disagree with in this type of interface are devilishly hard to recover from.
For example, spam detection can make two types of errors.
A false negative will place spam in a user's inbox, but that isn't a big issue because the decision and the error are visible and easy to fix.
A false positive is more problematic, as the user will not even know of the message unless they check their spam folder: the choice and the error are invisible, with the potential to wreak havoc.\footnote{Both Glassman and Kummerfeld have experienced further difficulties with spam detection and mail forwarding. When messages are labeled as spam by the forwarding account, they are not forwarded and so while there is a unified inbox, there is not a unified spam-box.
We have been unnecessarily stressed by messages silently and erroneously moved to our spam folders, as well as embarrassed by the messages we find there, unresponded to, now that we've had enough painful experiences to ingrain the habit of checking it regularly.}
Another example is summarization, where there can be errors of omission, e.g., leaving out information that a judge would ordinarily consider to be critical context when determining a convicted defendant's punishment. 
There can also be errors of misrepresentation: producing a shorter text for which, given the task at hand, the semantics shift beyond what would be acceptable, \textit{if the user noticed}.\footnote{Legal language is particularly tricky---for example, using the verbs ``will'' instead of ``shall'' or ``agree to assign'' instead of ``hereby assign'' can completely change how a legal professional in the US court system will interpret the semantics of the sentence, but even a fine-tuned foundation model may not robustly respect the semantic differences of what otherwise would be synonyms with no semantic distinctions.} 
The AI may also make more objective errors that, if not caught by the system, can be hard for the human to notice. For example, the errors introduced by LLM confabulation
within a summary look \textit{plausible at a glance} by definition.
Critically, undetected errors have \textit{both an immediate cost}---the user being insufficiently informed or inadvertently misled---\textit{and a long-term cost}, as the user who did not notice these errors cannot flag the output as incorrect for further model training.

\paragraph{Choices that are visible, but hard to notice}
A brief local change outside the narrow focus of a users' attention may go completely unnoticed.
The classic example of this form of inattentional blindness is people not seeing a man in a gorilla suit walking among students passing a ball when focused on a task that involves watching the students and the ball~\cite{inattentional}.
Inattentional blindness can be caused by limited cognitive resources, a target object's lack of salience, and the limitations of memory---\textit{a kind of seeing without noticing}~\cite{inattentional}.

\paragraph{Choices that are visible and noticed, but hard to understand}
Even in cases when users are aware that a change occurred, they may not understand the nature of the change.
For example, a global change to the layout of a cloud of points could play out in the interface without the user's ability to notice what actually changed; this is because noticing what changed would require having memory sufficient to capture the before state (which is now gone) and comparing it mentally to the current state.

\paragraph{Choices that are not recognized as a choice}
Sometimes choices go unnoticed because they are implicit. Perhaps we do not recognize a choice because it is presented as if it is inevitable or simply the truth rather than a choice of the designer or AI, making it an implicit choice. %example?
Other times, we do not recognize a choice because the option chosen for us is consistently chosen. This latter phenomenon shows up in a variety of different places: For example, in \citet{dowparallel}'s seminal work on parallel prototyping, participants commented on differences between analogous components\footnote{Components can only be determined to be analogous given a structural mapping. \citet{gentner1997structure}'s work concludes that participants implicitly look for potential structural mappings between objects, in order to find analogous components and note their differences.} across two prototypes (also called alignable differences) and neglected to comment on choices that were consistent across the two, either because (1) the participant did not think the designer was interested in feedback on that choice they made consistently across the two prototype designs or (2) the participants' attention was drawn to the differences and they did not notice or cognitive engage much with the consistent choices.\footnote{\citet{gentner1997structure}'s work found that participants identify more numerous and nuanced alignable differences between similar objects (that have more obvious structural mappings) than differences between less similar objects (that have less obvious structural mappings). It is unclear from that work whether (1) human cognition preferentially attends to alignable differences or (2) alignable differences are cognitively easier for us to compute (or both). If human cognition preferentially attends to alignable difference, it could be explained by the need to distinguish between edible and poisonous versions of the same type of plant, e.g. berries, during our evolution~\cite{gentner1997structure}.} Similarly, Variation Theory~\cite{marton} points out that not experiencing variation over a variable value can render that variable either unnoticed or undiscernable\footnote{The simple example Marton provides is that if you have no concept of color, you have to experience more than one color (variation in color) to discern the concept of color, at which point you can also recognize distinct points in color space.}; in the language of choices, this implies that not experiencing different choices at a particular choice point can render the choice cognitively invisible.

\paragraph{Choices that are visible, noticed, and understandable, but users choose not to make the effort to consider}
Finally, sometimes we know the choices are there but we choose not to cognitively engage much, like most people scrolling through yet another update to the terms and conditions.\footnote{And sometimes our habitual physical behaviors kick in before our cognitive engagement can, clicking on the button in the habitual location before we realize we have not actually thought about the information the button pertained to or even read the button label to confirm that the button was the one we habitually click!} This challenge is closely related to the challenge of fomenting and/or lowering barriers to cognitive engagement in AI outputs that have been acknowledged elsewhere, e.g., in a recent DARPA program call for systems that add friction in order to encourage users to engage with AI generated output more~\cite{darpa-fact}.

Cognitive engagement is a function of conscious and unconscious factors, but users cannot cognitively engage with something they do not notice. 
And when AI choices go unnoticed, they cannot be judged on whether they are objectively correct, contextually appropriate, or subjectively preferable.

\subsection{Insufficient context to judge (AI) choices}
\label{sec:insufficientcontext}
In order to judge an AI choice, the user needs access to sufficient context.
To be sufficient, this context must be enough for the user to come up with their own well-informed opinion in parallel to the AI.
An even stronger requirement would be that the context is only sufficient when any additional context would not change the user's choice.
\textit{Note that this is distinct from explanations of the AI's choice, as in AI explainability research.}
 
Making this judgement well can be hard for users, even when inattentional blindness is not at play, and we argue that many AI-powered features insufficiently support users in these tasks. This may be due to them (1) not being explicitly acknowledged in prior guidelines (and therefore more likely to be missed during the design and evaluation process), (2) being objectively difficult to design for, and/or (3) user resistance to slowing down when leveraging AI assistance, especially in low-stakes situations. 

For example, they need to read the entire summarized document as well as the summary to truly know whether they think, given their context and goals, the summary is appropriate. Interfaces that show the choice without all the relevant context risk either (1) the user confidently making a choice (while believing they have sufficient context) that does not reflect what they would have wanted if they had actually had all the context or (2) the user, knowing they do not have all the relevant context, being reduced to making a `gut' decision they may feel unsure about.

\section{AI-Resilient Interfaces}

Relative to prior human-AI interaction design guidelines and usability heuristics, including old classics like  ~\citet{gulfbook} and new canon like \citet{amershi2019guidelines}, the primary distinguishing goal of AI-resilient interfaces is to help users \emph{recognize} (1) objectively wrong AI choices, (2) contextually inappropriate AI choices, and (3) AI choices they subjectively dislike. So that users can make use of those previously described affordances that ``support efficient dismissal'' and ``support efficient correction --- [making] it easy to edit, refine, or recover when the AI system is wrong''~\cite{amershi2019guidelines}, because they know  it is necessary or desirable to traverse a gulf of execution~\cite{gulfbook} to recover from an AI choice. Especially in situations where the users' private context dominates their judgements. Even when their notion of what they want is evolving, possibly but not necessarily in response to the AI-resilient interface's features and affordances. Meeting these design goals should increase the safety, utility, and usability of the interface.

To make an AI-powered interface more AI-resilient, a designer may need to modify both (1) how AI choices are made visible to users, explicitly or implicitly, to support noticing and (2) providing sufficient information within the interface for the user to appropriately judge the correctness, appropriateness, and their subjective preferences over those AI choices.
This may make the interface more cognitively demanding and/or less traditionally usable, e.g., through less acceleration\footnote{This is referring to the acceleration of programmers described in ~\citet{groundedCopilot}} that can accumulate less thoughtful choices that become more painful to deal with later, e.g.,~\cite{priyanCopilotLBW}, but in both subjectively and objectively high-stakes situations, users may accept or even desire this if it makes them more resilient to AI choices they dislike or regard as wrong.

This AI-resilience may actually accelerate the user's intent formation, revision, and refinement---without driving it to a place that is globally worse than if the user had more slowly iterated on their intent without interface support. For example, in accordance with the Nielsen’s usability heuristic \emph{Recognition over Recall}, seeing alternative AI choices (or at least sufficient information scent about alternative choices) may allow users to confidently navigate between choices that serve them best in the moment---\textit{or discern entire new dimensions of alternatives of their own imagining} just based on the contrasts across multiple AI choices or between the AI suggestions and their own mental model of what they want, as implied by Variation Theory~\cite{marton}.

Given the challenges described previously, meeting each design goal depends on the context of the problem being solved. 
We will start with an example of an AI-resilient interface, Grammar-Preserving Text Saliency Modulation~\cite{gptsm}, to show concretely how it achieves the design goals for reading and skimming documents, and then discuss more generic tactics, with tradeoffs.

\subsection{A Concrete Example: An AI-Resilient Interface Alternative to AI Summarization}
\label{sec:gptsm}

\subsubsection{Specific AI-Resilience Needs}
AI-powered systems can, however, make choices that are hard to notice. For example, when performing summarization, generative AI may (1) omit context critical to correctly understanding an output, (2) impute information that is statistically likely given the training data (and therefore likely plausible to the human consumer) but not present in the original document(s) or (3) subtly or significantly misrepresent the original document(s)' meaning through alternative word choices or simplification. All three AI choices are tedious and memory-intensive for humans to check, negating a lot of the value of a summary, and delegating this check to another AI, which can be confidently wrong, does not solve the issue.

\begin{figure}
\fbox{\parbox{6cm}{The world is \textcolor{lightgray}{at present} accumulating carbon dioxide \textcolor{gray}{in the atmosphere} from two \textcolor{lightgray}{well-known} sources: \textcolor{gray}{the combustion of} fossil fuels and deforestation.}}
\caption{\label{fig-gptsm}GP-TSM~\cite{gptsm} output, with multiple levels of text opacity revealing levels of AI-predicted semantic criticality, while keeping all original text (context) legible.}
\end{figure}

\subsubsection{AI-Resilient Approach}
The goal of grammar-preserving text salience modulation (GP-TSM)~\cite{gptsm} was AI-resilient single-document summarization.
The approach used a large language model to perform recursive sentence compression, i.e., recursively identifying additional words that can be removed from the text while (a) retaining the core meaning and (b) leaving a grammatical result for ease of reading.
Each iteration of compression corresponds to a different level of text opacity.
Figure~\ref{fig-gptsm} shows an example output, which shows the most concise summary (in black) and a skimmable sense of varying levels of detail (in varying shades of gray).
Even the lightest text is still legible so readers have all the context necessary to decide whether they agree with the de-emphasis of various words, and if they disagree, they can nearly effortlessly cognitively recover by reading more without taking a physical action. As described in~\cite{gptsm}, prior versions allowed users to hide text below a user-chosen threshold of salience, but this removed its AI-resiliency, since users cannot notice hidden words they do not think should be hidden.

In terms of our AI-resilience design goals, this system (1) allows users to continuously choose what level of summarization at which to read, (2) visualizes the decisions made by the model using text saliency, and (3) retains readability of de-emphasized text with minimal effort. User studies indicate that readers could answer reading comprehension questions more accurately in less time when reading text rendered with GP-TSM, relative to both prior art in text salience modulation and normal, constant-salience text, and, because they did not read a summary, which isn't an AI-resilient technology, they did not suffer from any AI choices they could not recognize and recover from.

\section{Relationship with Trust and Design Principles} \label{sec:principles}
% \todo{add co-audit papers}
% \todo{add this thought, fleshed out: "With this definition of AI-resilient interfaces, we 
% at least partially sidestepping long-standing difficult research problems, e.g., over/underreliance due to over/under trust."}

% \todo{Sensible AI, Kaur paper}
% \cite{kaur-sensible-AI}

%  \todo{add reference to iceberg sensemaking paper when discussing context to make decisions --- Elena}
% \todo{add "Interpretable Outputs: Criteria for Machine Learning in the Humanities" https://www.digitalhumanities.org/dhq/vol/15/2/000555/000555.html -- Elena}

% \todo{Related work: connect to Oster's argument about economics and rules vs personalized trade-offs in Expecting Better (or better yet, elsewhere)}
%\todo{add connection to antagonism}

%Designing for AI-resilience sidesteps or reduces some issues of trust, but not all.

%\subsection{Trust, Interpretability, Transparency, and Simulability, and Appropriate Reliance}
\subsection{Trust}
When interacting with AI systems, users typically need to develop and maintain an accurate mental model of where it is and is not appropriate to trust the system~\cite{bansal2019beyond}.
But a less precise sense of trust is needed with an AI-resilient system because the user can more easily identify and recover from the model's choices they disagree with.
AI-resilient systems may also make it easier for users to refine their sense of trust, because the interface design should help users see when the model makes choices (in)congruent with their context, preferences, and goals.

% This is similar to, but not the same as, notions of algorithmic transparency, simulability, and interpretability.
% The similarity comes in the goals, but those areas of study typically focus on the process inside the AI model itself.
% %The ideas we consider only draw out more information from the final step of the model, e.g., indicating the probability assigned to each possible outcome.
% Other examples of related issues that may be mitigated include 
% algorithm aversion after observing the algorithm err~\cite{dietvorst2015algorithm} and over-reliance.

\subsection{Existing Design Principles}
A range of researchers, companies, organizations, and governments have proposed principles for responsible development of systems that incorporate AI.
As previously stated, (1) noticing AI choices and (2) having enough context to judge whether or not they agree with each AI choice are pre-requisites to using the affordances recommended by popular human-AI interaction design guidelines~\cite{amershi2019guidelines}, specifically \textit{support efficient dismissal} and \textit{support efficient correction}. 
\citet{ai-principles} proposed several principles which relate to AI-resilience: \textit{Autonomy: The Power to Decide}, \textit{Justice: Promoting Prosperity, Preserving Solidarity, Avoiding Unfairness}, and \textit{Explicability: Enabling the Other Principles through Intelligibility and Accountability}.
Autonomy depends on being able to notice the AI choice in order to make decisions about it, and \emph{meaningful autonomy} rather than a faux-autonomy requires enough context to make that decision well rather than arbitrarily.
The avoiding unfairness component of justice could be assisted by AI-resilience, as users can more easily identify decisions that are not suitable for them in their context.
Intelligibility is also well served by our notion of noticing and having the context to judge.

\section{AI Resilience Audits of Current Approaches}

Now that we have defined AI resilience, it is natural to ask how many existing approaches satisfy it?
Some tasks may not be amenable to this paradigm, or may require significant changes to the standard form of input and output.
In this section, we consider a range of tasks that have AI as a core part of common solutions.
For each task, we consider whether the dominant approach to supporting users is AI-resilient.
In a few cases, we also describe a specific interface and how AI resilient it is.

\subsection{Automated Document Summarization}

In Section~\ref{sec:gptsm}, we described an approach for building an AI resilient version of automated document summarization.
Here, we consider the typical formulation of the task.
In either single document or multi-document summarization, the input is a large volume of text and the output is a much shorter piece of text that is intended to contain the same core meaning.

When reading a summary, the user does not know where each part is derived from, why they were chosen, or what alternatives were considered.
In the abstractive case, where the summary is newly generated text, it may contain entirely incorrect content.
Even in the extractive case, where the summary is composed of parts of the input, there is the possibility of misleading the user by leaving out important context or putting content together in ways that suggest invalid conclusions.
The standard fallback option for a user is to read the entire document or document collection, with a memory approaching super-humanity levels.
This does not meet any of our design goals for AI resilient interfaces, as the user does not see the choices made by the AI and the only means of judging and recovery is onerous.

\subsection{Recommender Systems}

The search examples from Section~\ref{sec:twins} involve an AI model that ranks pages in response to the user query.
This is closely related to recommender systems, which rank items, often with the goal of being personalized for each user.
Unlike search, where there is a query driving the ranking, here the ranking is typically based on prior ratings provided by the user.
However, while the user data does lead to personalized results, the algorithm is the same across all users.

In this setting, the user either does not see the options that were not recommended, or they may be far enough down the list of recommendations that they are essentially invisible.
The impact of user actions is also opaque, with no indication of how a user's ratings are shaping the recommendations they receive.

\citet{user-control-rec} created one example of work that gives users some awareness and agency.
Specifically, the system provides users with some controls that directly change the equations in the algorithm.
Critically, the impact of those changes is shown to users by showing movie recommendations that were added and removed in response to the user action.
Their user study also showed that participants significantly preferred the results when given awareness and agency, demonstrating the value of methods that increase AI resilience.

\subsection{Unsupervised Pattern Identification}

Vector representations, i.e., embeddings, are the core of modern AI methods across many subfields of AI, including for audio, visual, written, and state data.\footnote{By state data, we intend to encompass systems that do planning and control.}
In a range of situations, it is helpful to use embeddings to identify patterns in data, e.g., characterizing a collection of documents, photos, or videos. 
A common approach to this task is to take the high dimensional vector space (100s to 1,000s of dimensions), identify clusters in the space, project down to two dimensions, and show the space as a scatter plot, with the ability to reveal, at the user's request, the item each dot represents, e.g., by hovering over the dots one by one.

The choices in the process of producing these representations are entirely opaque to the user.
While all the data is accessible to users in this format, it is not practical to see much of it, let alone remember item contents in order to recognize relationships that the user cares about more or less than the AI that generated the embedding space.
The most computationally prioritized pattern(s) dominate the visualization, making it difficult to see more subtle global and local patterns.
There may also be patterns that are not captured by the vector space, which could involve items that are spread all over the space.
The hover affordance does not resolve this, as users can see at most a handful of items at a time and need to remember what they have seen previously to identify patterns.
Popular methods, such as t-SNE~\cite{tsne}, also involve a range of configuration parameters that can radically change the appearance of the space~\cite{wattenberg2016how}.
For these reasons, despite being superficially AI resilient (since all data is present and theoretically accessible via hovering or selecting regions~\cite{explainAndTest}), these visualizations are not functionally AI resilient.

\subsection{AI-Assisted Writing Interfaces}

AI-assistance for writing has dramatically changed in recent years.
While spelling and grammar correction and text prediction have existed in some form for decades, newer larger language model technologies can go considerably further in shaping the writing process.
The most common set-up involves providing one or a small sampling of potential alternative continuations of the writer's current text, e.g.,~\cite{elephant, metaphoria}.

Even when a few options are shown, prior work has shown that if the options are not carefully selected, people focus on certain choices while not realizing other options were available too~\cite{arnold2020predictive}.
\citet{ws-neurips-hegm:22:lm} argue that many more samples are needed to accurately build intuition about the behavior of LLMs.
Since the space of possible text continuations is exponential, it is not feasible to show them all. Designing how and what to render in order to provide more utility to users while not overwhelming them is an open research direction.

\subsection{Machine Translation}

Starting from the 1940s~\cite{weaver}, the idea of automatic translation between human languages has been a major area in AI.\footnote{In fact, work on translation pre-dates the term Artificial Intelligence, which was coined in 1956~\cite{Moor_2006}.}
Today, widely used systems exist for translation between many language pairs and multiple modalities.

The AI-resilience of these systems is extremely variable.
The most critical factor is the language skills of the user.
If they have no knowledge at all of the other language, they will struggle to identify any mistake, while a somewhat knowledgeable user may notice egregious errors, and an experienced one may pick up on subtle model choices.

AI-resilience of the translation interface will also vary depending on the modality of the model and the setting in which translation occurs.
For speech-to-speech translation, it is not possible to statically view the two sides of the translation in order to recognize errors, while for text-to-text translation, alignment and glossing methods could help reveal misunderstandings and/or mistranslations.
In a conversational setting, the AI choices that result in translated content is transient, making user noticing and judgement of errors more difficult.\footnote{Though at the same time, the other entity in the conversation may be able to assist in identifying and rectifying errors. We use entity rather than person since it is forseeable that the other interlocutor may be an AI model.} 
In contrast, in an offline setting, there is time to examine and resolve issues over a longer time-scale.

\subsection{Code Generation for Programmers}

One rapid uptake of large language models is for programming assistance.
Developers either receive spontaneous code suggestions from a model, or they write a description of the code they need and the model generates a solution.
In both cases, the results can often be used verbatim.
This is being used both for general purpose languages, e.g., with GitHub Copilot, and in specialized languages like spreadsheet macros, e.g., Google Sheets.

This may seem to satisfy many aspects of AI-resilience.
The developer is always shown the choice made by the model (i.e., what code to write), they have full access to their local context (i.e., their local code, which may be shared with the AI as common ground, and their not yet explicit programming goals and preferences) needed to judge the code's usefulness, and they are able to choose whether or not to use it.
However, while they are shown the AI's chosen generated code, they may not cognitively engage with it.
If, at a quick glance, the code seems plausibly correct, the developer may accept the suggestion without much thought, as programmers in~\cite{priyanCopilotLBW} may have been doing when they performed worse on programming problems than without AI assistance.
Even if they look closely, a developer may accept a suggestion without noticing it contains a bug and/or does not do exactly what they wanted. 
And even if it is doing the task correctly, it may not be the most efficient approach, a fact the developer may not realise without putting in the effort to identify alternatives. 

In some cases, the programmer can see other options the model generated, in a dropdown list, but the programmer may be left to their own devices to sort through the alternatives and compare and contrast each one manually without any precomputed commonalities and differences rendered in the interface, unlike the interface support available in prior systems for viewing tens, hundreds, or thousands related examples of text (like \citet{ws-neurips-hegm:22:lm} and~\cite{mesotext}) and code (like OverCode~\cite{overcode}, ExampleStack~\cite{exampleStack}, Examplore~\cite{examplore}, and ParaLib~\cite{paralib}).

\subsection{Discussion}

AI-resilience is not a panacea nor easy to implement. As discussed in Section~\ref{sec:principles}, it only addresses some of the AI principles and issues in systems today.

The analysis above also reveals a (hopefully productive) tension between users' cognitive load and the AI-resilience their interface affords them.
Providing the additional context needed to notice and judge AI choices often incurs friction in the user experience.
In small amounts, that friction can be positive, but at some point the cost may be too high, leading users to stop using an AI-resilient system altogether.
Achieving that balance is a challenge, and one that requires challenging strongly held assumptions about the formulation of our tasks.
Even if an AI-resilient interface cannot be achieved, the exercise of trying to develop one may be enlightening, revealing issues in system design that are missed when we are following the established patterns of past work. 

We may be reluctant to ``burden'' users with what would be heuristically evaluated as too much information if we cannot find a way, like GP-TSM did, to reify that information in a way that improves rather than detracts from user experience and performance. We may also underestimate how much revealing structure within the variation over a larger amount of data may support more confident sensemaking and reduce its cognitive burden.

We have also generally assumed that the user knows best.
What if the AI choices are better aligned with the user's ultimate refined understanding of what they want but poorly aligned with what the user \emph{currently} wants?
In other words, is it possible for the user \textit{not} to be ``always right''?
There are several scenarios to consider: First, the user could be considered, by definition, always right (in that moment) about what serves them best, and the interaction with the AI-resilient system may help them eventually iterate their way to a set of beliefs about what they want and need from the AI interface that serves them better.
That seems like a successful human-AI interaction, even if the system is ``only'' helping the user converge on those final beliefs through its legible AI choices and alternatives and quickly reflecting the user's current beliefs so that the user can understand how those beliefs in that moment do and do not need to be updated.
In contrast, if the system is less deferential to the user and stands its ground, more `paternalistically' or antagonistically~\cite{antagonistic_ai}, how does that affect the rate at which users arrive at their final beliefs about what is best for them, and which final beliefs they arrive at?
How does that play out when the system is confidently wrong in some objective sense about the user's context but manages to obscure that from the user?
The answers to these questions depend on many factors, including the task, the user, and the time available. While no one-size-fits-all AI-resilient interface affordances may crystallize in the next decade, we hope the concerns and tactics described here augment the human-AI design guidelines already available in a way that is critical to AI safety, utility, and usability.

\begin{acks}
This material is based upon work supported by the National Science Foundation under Grant No. IIS-2107391. This work was also supported by the Sloan Research Fellowship. Much gratitude to the expertise and effort of Ian Arawjo, who helped brainstorm and evaluate early versions of Grammar-Preserving Text Saliency Modulation. 
\end{acks}

% \section*{Contributions Statement}
% ELG proposed the core initial idea, which both authors then developed together, with input from the Variation Lab.
% The Introduction, Search Example, Design subgaols, and Related Work, were primarily written by ELG.
% The Corpus Example, and Audits were primarily written by JKK.
% All other sections were jointly written.
% Both authors edited and gave feedback on all sections.

\bibliographystyle{ACM-Reference-Format}
\bibliography{sample-base}

%\appendix

\end{document}